\documentclass[aps,superscriptaddress,preprintnumbers,nofootinbib,floatfix]{revtex4}
\usepackage{amssymb}
\usepackage{bm}
\usepackage{graphicx}

\begin{document}

%===================  TITLE, AUTHORS, AFFILIATIONS ===================
%\newcommand*{\xxx}{bla, bla}\affiliation{\xxx}
\newcommand*{\Dubna}{
	Bogoliubov Laboratory of Theoretical Physics, JINR, 141980 Dubna, 
	Russia}\affiliation{\Dubna}
\newcommand*{\UConn}{
        Department of Physics, 
        University of Connecticut, Storrs, CT 06269, USA}\affiliation{\UConn}
\newcommand*{\Praha}{
	Institute of Physics of the AS CR, 
	Na Slovance 2, CZ-182 21 Prague 8, Czech Rep.}\affiliation{\Praha}
\title{Quark intrinsic motion and the link between TMDs and PDFs in 
covariant approach
\footnote{Contribution to the Proceedings of XIII Workshop on High Energy %
Spin Physics DSPIN-09, Dubna, Russia, September 1-5, 2009}}

\author{A.~V.~Efremov}\affiliation{\Dubna}
\author{P.~Schweitzer}\affiliation{\UConn}
\author{O.~V.~Teryaev}\affiliation{\Dubna}
\author{P.~Zavada}\affiliation{\Praha}

\begin{abstract}
\noindent
The relations between TMDs and PDFs are obtained from the symmetry %
requirement: relativistic covariance combined with %
rotationally symmetric parton motion in the nucleon rest frame. This %
requirement is applied in the covariant parton model. Using the usual %
PDFs as an input, we are obtaining predictions for some polarized and %
unpolarized TMDs.  
\end{abstract}

\maketitle

The transverse momentum dependent parton distribution functions (TMDs) \cite%
{tmds,Mulders:1995dh} open the new way to more complete understanding of the
quark-gluon structure of the nucleon. We studied this topic in our recent
papers \cite{Efremov:2009ze,Avakian:2009jt,Zavada:2009sk}. We have shown,
that requirements of symmetry (relativistic covariance combined with
rotationally symmetric parton motion in the nucleon rest frame) applied in
the covariant parton model imply the relation between integrated unpolarized
distribution function and its unintegrated counterpart. Obtained results are
shortly discussed in the first part. Second part is devoted to the
discussion of analogous relation valid for polarized distribution functions.

\textbf{Unpolarized distribution function}

\begin{figure}[tbp]
\includegraphics[width=12cm]{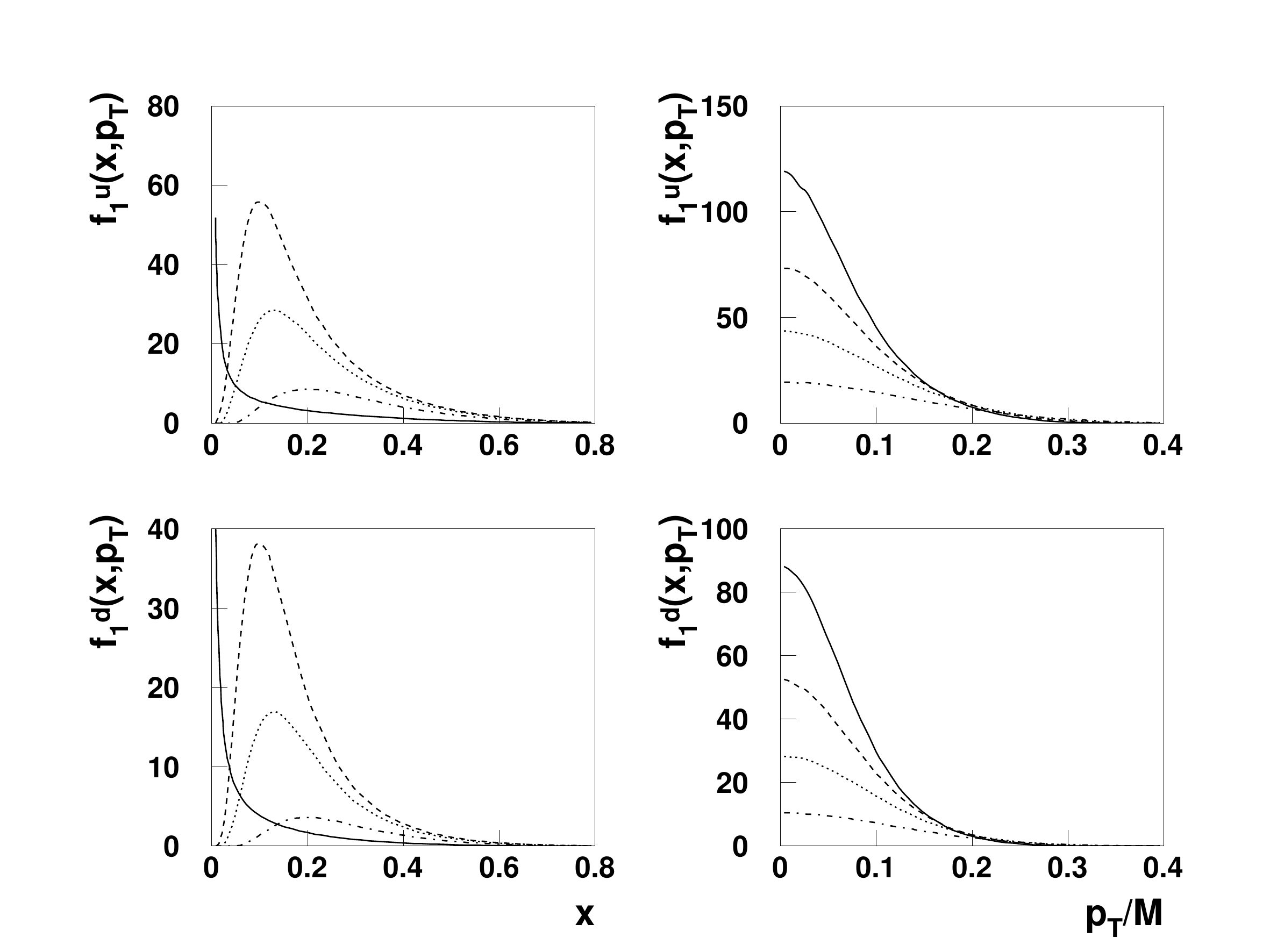}
\caption{Transverse momentum dependent unpolarized distribution functions
for $u$ (\textit{upper figures}) and $d-$quarks (\textit{lower figures}). 
\textbf{Left part}: dependence on $x$ for $p_{T}/M=0.10,0.13,0.20$ is
indicated by dash, dotted and dash-dot curves; solid curve correspods to the
integrated distribution $f_{1}^{q}(x)$. \textbf{Right part}: dependence on $%
p_{T}/M$ $\ $for $x=0.15,0.18,0.22,0.30$ is indicated by solid, dash, dotted
and dash-dot curves.}
\label{ff1}
\end{figure}

\begin{figure}[tbp]
\includegraphics[width=12cm]{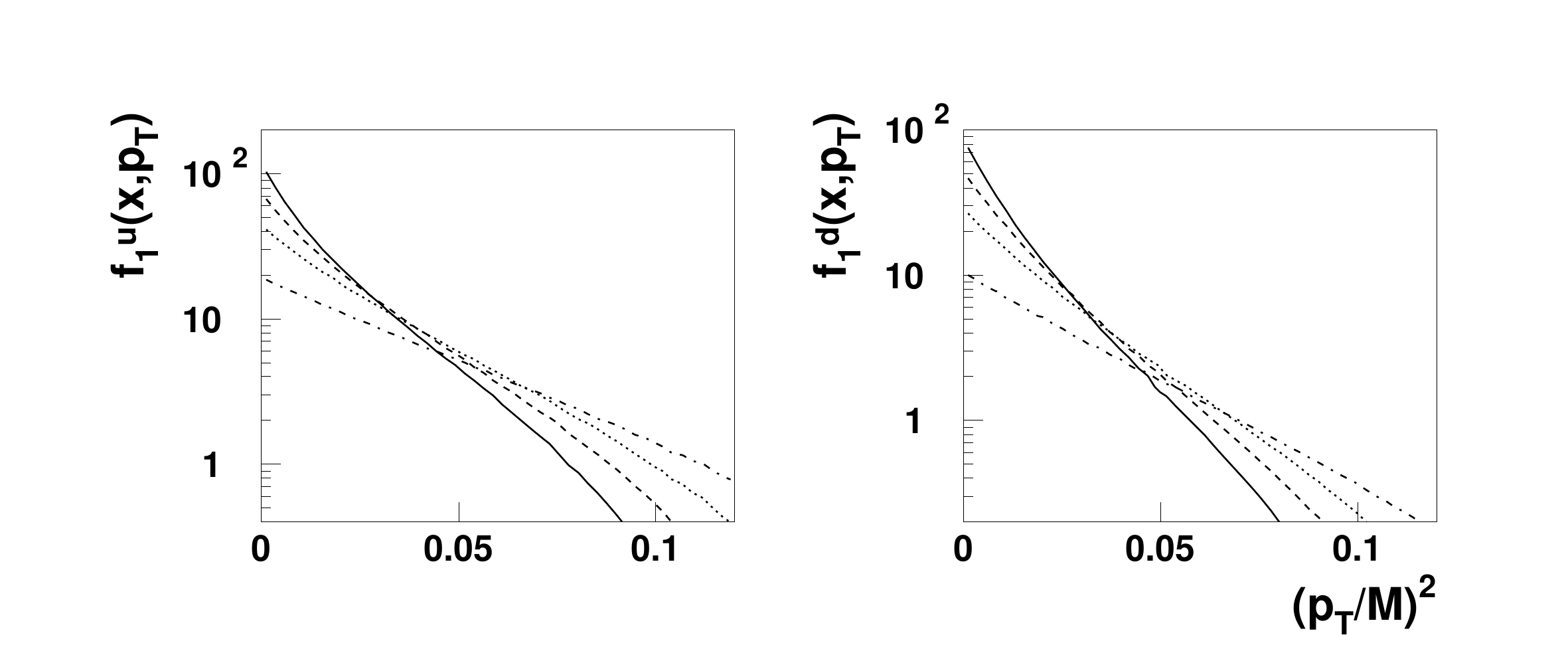}
\caption{Transverse momentum dependent unpolarized distribution functions
for $u$ and $d-$quarks. Dependence on $\left( p_{T}/M\right) ^{2}$ $\ $for $%
x=0.15,0.18,0.22,0.30$ is indicated by solid, dash, dotted and dash-dot
curves.}
\label{ff2}
\end{figure}

In the covariant parton model we showed \cite{Zavada:2007ww}, that the
parton distribution function $f_{1}^{q}(x)$ generated by the 3D distribution 
$G_{q}$ of quarks reads:%
\begin{equation}
f_{1}^{q}(x)=Mx\int G_{q}(p_{0})\delta \left( \frac{p_{0}+p_{1}}{M}-x\right) 
\frac{dp_{1}d^{2}\mathbf{p}_{T}}{p_{0}}  \label{q1}
\end{equation}%
and that this integral can be inverted%
\begin{equation}
G_{q}\left( \frac{M}{2}x\right) =-\frac{1}{\pi M^{3}}\left( \frac{%
f_{1}^{q}(x)}{x}\right) ^{\prime }.  \label{q10}
\end{equation}%
Further, due to rotational symmetry of the distribution $G_{q}$ in the
nucleon rest frame, the following relations for unintegrated distribution
were obtained \cite{Zavada:2009sk}:%
\begin{equation}
f_{1}^{q}(x,\mathbf{p}_{T})=MG_{q}\left( \frac{M}{2}\xi \right) .  \label{q9}
\end{equation}%
After inserting from Eq. (\ref{q10}) we get relation between unintegrated
distribution and its integrated counterpart:%
\begin{equation}
f_{1}^{q}(x,\mathbf{p}_{T})=-\frac{1}{\pi M^{2}}\left( \frac{f_{1}^{q}(\xi )%
}{\xi }\right) ^{\prime };\qquad \xi =x\left( 1+\left( \frac{p_{T}}{Mx}%
\right) ^{2}\right) .  \label{q11}
\end{equation}%
Now, using some input distributions $f_{1}^{q}(x)$ one can calculate
transverse momentum distribution functions $f_{1}^{q}(x,\mathbf{p}_{T})$. As
the input we used the standard PDF parameterization \cite{Martin:2004dh} (LO
at the scale $4GeV^{2}$).\ In Fig. \ref{ff1} we have results obtained from
relation (\ref{q11}) for $u$ and $d-$quarks. 
The right part of this figure is shown again, but in different scale in Fig %
\ref{ff2}. 
One can observe the following:

\textit{i)} For fixed $x$ the corresponding $p_{T}-$ distributions are very
close to the Gaussian distributions%
\begin{equation}
f_{1}^{q}(x,p_{T})\propto \exp \left( -\frac{p_{T}^{2}}{\left\langle
p_{T}^{2}\right\rangle }\right) .  \label{c20}
\end{equation}

\textit{ii)} The width $\left\langle p_{T}^{2}\right\rangle =\left\langle
p_{T}^{2}(x)\right\rangle $ depends on $x$. This result corresponds to the
fact, that in our approach, due to rotational symmetry, the parameters $x$
and $p_{T}$ are not independent.

\textit{iii)} Figures suggest the typical values of transversal momenta, $%
\left\langle p_{T}^{2}\right\rangle \approx 0.01GeV^{2}$ or $\left\langle
p_{T}\right\rangle \approx 0.1GeV$. These values correspond to the estimates
based the analysis of the experimental data on structure function $%
F_{2}(x,Q^{2})$ \cite{Zavada:2009sk}. They are substantially lower, than the
values $\left\langle p_{T}^{2}\right\rangle \approx 0.25GeV^{2}$ or $%
\left\langle p_{T}\right\rangle \approx 0.44GeV$ following e.g. from the
analysis of data on the Cahn effect \cite{Anselmino:2005nn} or HERMES data 
\cite{Collins:2005ie}. At the same time the fact, that the shape of obtained 
$p_{T}-$ distributions (for fixed $x$) is close to the Gaussian, is
remarkable. In fact, the Gaussian shape is supported by phenomenology.

\textbf{Polarized distribution functions}

Relation between the distribution $g_{1}^{q}(x)$ and its unintegrated
counterpart can be obtained in a similar way, however in general the
calculation with polarized structure functions is slightly more complicated.
First let us remind procedure for obtaining structure functions $g_{1},g_{2}$
from starting distribution functions $G^{\pm }$ defined in \cite%
{Zavada:2007ww},\ Sec. 2, see also the footnote there. In fact the auxiliary
functions $G_{P},G_{S}$ are obtained in appendix of the paper \cite{zav1}.
If \ we assume that $Q^{2}$ $\gg 4M^{2}x^{2}$, then the approximations%
\begin{equation}
\left\vert \mathbf{q}\right\vert \approx \nu ,\qquad \frac{pq}{Pq}\approx 
\frac{p_{0}+p_{1}}{M}  \label{m1}
\end{equation}%
are valid and the equations (A1),(A2) can be with the use of (A3),(A4)
rewritten as%
\begin{equation}
G_{X}=\int \Delta G\left( p_{0}\right) w_{X}\delta \left( \frac{p_{0}+p_{1}}{%
M}-x\right) \frac{dp_{1}d^{2}\mathbf{p}_{T}}{p_{0}},\qquad X=P,S  \label{m2}
\end{equation}%
where%
\begin{eqnarray}
w_{P} &=&-\frac{m}{2M^{2}\nu }\frac{-p_{1}\cos \omega +p_{T}\cos \varphi
\sin \omega }{p_{0}+m}  \label{m3} \\
&&\times \left( 1+\frac{1}{m}\left( p_{0}-\frac{-p_{1}-\left( -p_{1}\cos
\omega +p_{T}\cos \varphi \sin \omega \right) \cos \omega }{\sin ^{2}\omega }%
\right) \right) ,  \nonumber
\end{eqnarray}%
\begin{eqnarray}
w_{S} &=&\frac{m}{2M\nu }\left( 1+\frac{-p_{1}\cos \omega +p_{T}\cos \varphi
\sin \omega }{p_{0}+m}\frac{1}{m}\right.  \label{m4} \\
&&\times \left. \left( -p_{1}\cos \omega +p_{T}\cos \varphi \sin \omega -%
\frac{-p_{1}-\left( -p_{1}\cos \omega +p_{T}\cos \varphi \sin \omega \right)
\cos \omega }{\sin ^{2}\omega }\cos \omega \right) \right) .  \nonumber
\end{eqnarray}%
Let us remark, that using the notation defined in \cite{Efremov:2009ze} we
can identify%
\begin{equation}
-\cos \omega =S_{L},\qquad \sin \omega =S_{T},\qquad p_{T}\sin \omega \cos
\varphi =\mathbf{p}_{T}\mathbf{S}_{T},  \label{m4b}
\end{equation}%
which appear in definition of the TMDs \cite{Mulders:1995dh}:%
\begin{equation}
\frac{1}{2}\mathrm{tr}\left[ \gamma ^{+}\gamma _{5}\phi ^{q}(x,\mathbf{p}%
_{T})\right] =S_{L}g_{1}^{q}(x,p_{T})+\frac{\mathbf{p}_{T}\mathbf{S}_{T}}{M}%
g_{1T}^{\bot q}(x,p_{T}).  \label{m4a}
\end{equation}%
Now, in analogy with Eq. (46) in \cite{zav1} we define (note that $%
Pq/qS=-M/\cos \omega $): 
\begin{equation}
w_{1}=M\nu \cdot w_{S}+\frac{M^{2}\nu }{\cos \omega }\cdot w_{P},\qquad
w_{2}=-\frac{M^{2}\nu }{\cos \omega }\cdot w_{P},  \label{m7}
\end{equation}%
which implies%
\begin{equation}
g_{k}^{q}=\int \Delta G\left( p_{0}\right) w_{k}\delta \left( \frac{%
p_{0}+p_{1}}{M}-x\right) \frac{dp_{1}d^{2}\mathbf{p}_{T}}{p_{0}},\qquad
k=1,2.  \label{m8}
\end{equation}%
From Eqs.(\ref{m3}),(\ref{m4}) and the definition (\ref{m7}) we obtain%
\begin{equation}
w_{1}=\frac{1}{2}\left( m+p_{1}\left( 1+\frac{p_{1}}{m+p_{0}}\right)
-p_{T}\tan \omega \left( 1+\frac{p_{1}}{m+p_{0}}\right) \allowbreak \cos
\varphi \right) ,  \label{m9}
\end{equation}%
Apparently, the terms proportional to $\cos \varphi $ disappear\ in the
integrals (\ref{m8}) and the remaining terms give structure functions $%
g_{1},g_{2}$ defined by Eqs. (15),(16) in \cite{Zavada:2007ww}.

Now, the integration over $p_{1}$ and further procedure can be done in a
similar way as for unpolarized distribution. First, to simplify calculation,
we assume $m\rightarrow 0.$ For $w_{1}$\ we get%
\begin{equation}
g_{1}^{q}(x)=\frac{1}{2}\int \Delta G_{q}(p_{0})\left( 1+\frac{p_{1}}{p_{0}}%
\right) \left( p_{1}-p_{T}\tan \omega \allowbreak \cos \varphi \right)
\delta \left( \frac{p_{0}+p_{1}}{M}-x\right) \frac{dp_{1}d^{2}\mathbf{p}_{T}%
}{p_{0}}.  \label{m12}
\end{equation}%
The $\delta -$function is modified as%
\begin{equation}
\delta \left( \frac{p_{0}+p_{1}}{M}-x\right) dp_{1}=\frac{\delta \left(
p_{1}-\tilde{p}_{1}\right) dp_{1}}{x/\tilde{p}_{0}},  \label{m13}
\end{equation}%
where 
\begin{equation}
\tilde{p}_{1}=\frac{Mx}{2}\left( 1-\left( \frac{p_{T}}{Mx}\right)
^{2}\right) ,\qquad \tilde{p}_{0}=\frac{Mx}{2}\left( 1+\left( \frac{p_{T}}{Mx%
}\right) ^{2}\right) .  \label{m14}
\end{equation}%
\begin{figure}[tbp]
\includegraphics[width=12cm]{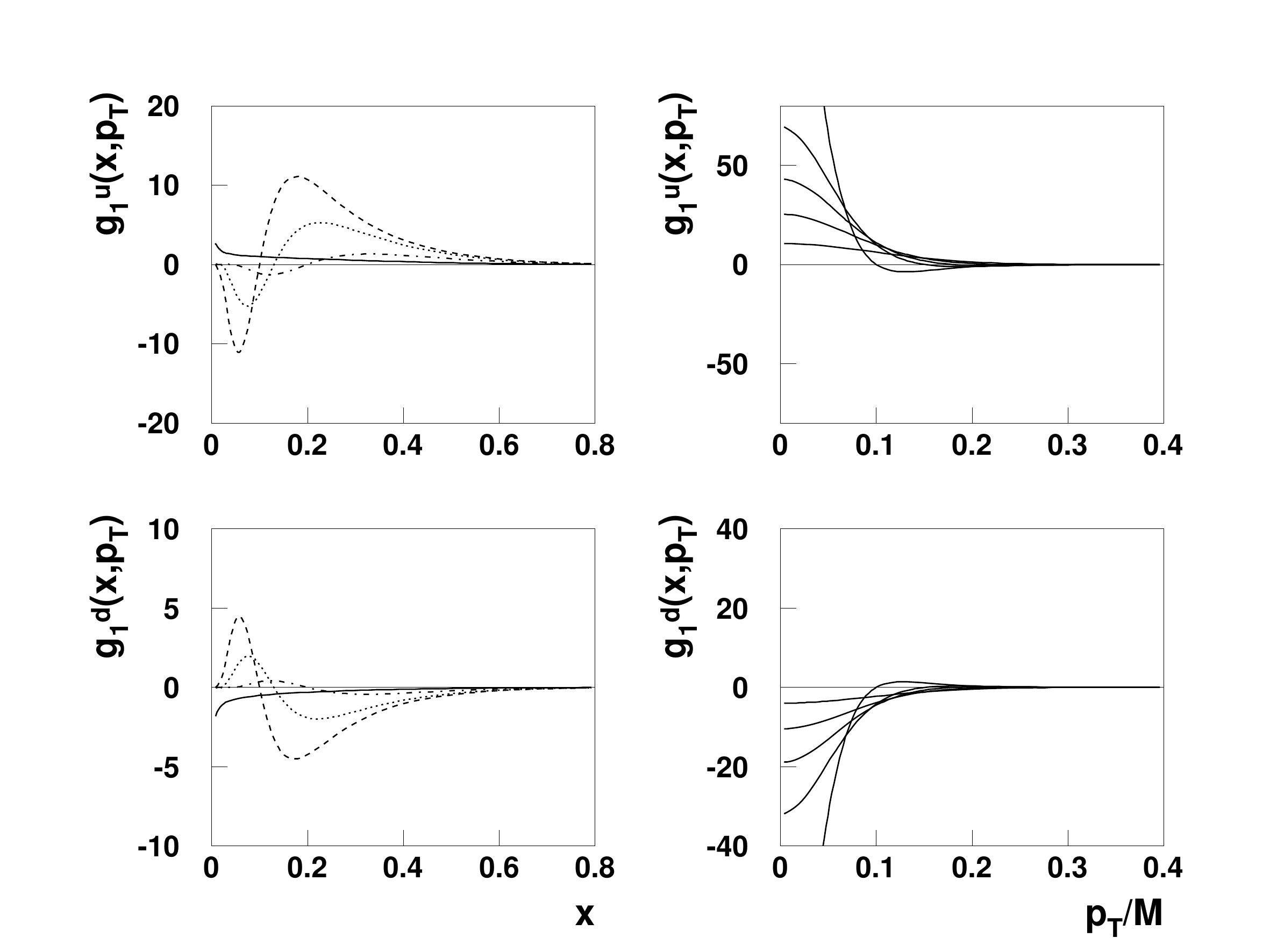}
\caption{Transverse momentum dependent polarized distribution functions for $%
u$ (\textit{upper figures}) and $d-$quarks (\textit{lower figures}). \textbf{%
Left part}: dependence on $x$ for $p_{T}/M=0.10,0.13,0.20$ is indicated by
dash, dotted and dash-dot curves; solid curve correspods to the integrated
distribution $g_{1}^{q}(x)$. \textbf{Right part}: dependence on $p_{T}/M$ $\ 
$for $x=0.10,0.15,0.18,0.22,0.30$ from top to down for $u-$quarks, and the
same symmetrically for $d-$quarks.}
\label{ff3}
\end{figure}
Modified $\delta -$function allows to simplify the integral

\begin{equation}
g_{1}^{q}(x)=\frac{1}{2}\int \Delta G_{q}(\tilde{p}_{0})\left( M\left(
2x-\xi \right) -2p_{T}\tan \omega \allowbreak \cos \varphi \right) \frac{%
d^{2}\mathbf{p}_{T}}{\xi },  \label{m15}
\end{equation}%
where%
\begin{equation}
\xi =x\left( 1+\left( \frac{p_{T}}{Mx}\right) ^{2}\right) .  \label{q8}
\end{equation}%
Now we define%
\begin{equation}
\Delta q(x,\mathbf{p}_{T})=\frac{1}{2}\Delta G_{q}\left( \frac{M\xi }{2}%
\right) \left( M\left( 2x-\xi \right) -2p_{T}\tan \omega \cos \varphi
\right) \frac{1}{\xi }.  \label{m16}
\end{equation}%
According to Eq. (40) in \cite{Zavada:2007ww} we have%
\begin{equation}
\Delta G_{q}\left( \frac{M\xi }{2}\right) =\frac{2}{\pi M^{3}\xi ^{2}}\left(
3g_{1}^{q}(\xi )+2\int_{\xi }^{1}\frac{g_{1}^{q}(y)}{y}dy-\xi \frac{d}{d\xi }%
g_{1}^{q}(\xi )\right) .  \label{m17}
\end{equation}%
After inserting to Eq. (\ref{m16}) one gets: 
\begin{eqnarray}
\Delta q(x,\mathbf{p}_{T}) &=&\frac{1}{\pi M^{2}\xi ^{3}}\left(
3g_{1}^{q}(\xi )+2\int_{\xi }^{1}\frac{g_{1}^{q}(y)}{y}dy-\xi \frac{d}{d\xi }%
g_{1}^{q}(\xi )\right)   \label{m18} \\
&&\times \left( 2x-\xi -2\frac{p_{T}}{M}\tan \omega \allowbreak \cos \varphi
\right) .  \nonumber
\end{eqnarray}%
This relation allows us to calculate the distribution $\Delta q(x,\mathbf{p}%
_{T})$ from a known input on $g_{1}^{q}(x)$. Further, it can be shown, that
using the notation defined in Eqs. (\ref{m4b}),(\ref{m4a}), our result reads%
\begin{equation}
-\cos \omega \cdot \Delta q(x,\mathbf{p}_{T})=S_{L}g_{1}^{q}(x,p_{T})+\frac{%
\mathbf{p}_{T}\mathbf{S}_{T}}{M}g_{1T}^{\bot q}(x,p_{T}),  \label{m19}
\end{equation}%
where%
\begin{equation}
g_{1}^{q}(x,p_{T})=\frac{2x-\xi }{\pi M^{2}\xi ^{3}}\left( 3g_{1}^{q}(\xi
)+2\int_{\xi }^{1}\frac{g_{1}^{q}(y)}{y}dy-\xi \frac{d}{d\xi }g_{1}^{q}(\xi
)\right) ,  \label{m20}
\end{equation}%
\begin{equation}
g_{1T}^{\bot q}(x,p_{T})=\frac{2}{\pi M^{2}\xi ^{3}}\left( 3g_{1}^{q}(\xi
)+2\int_{\xi }^{1}\frac{g_{1}^{q}(y)}{y}dy-\xi \frac{d}{d\xi }g_{1}^{q}(\xi
)\right) .  \label{m21}
\end{equation}%
Apparently, both functions are related in our approach:%
\begin{equation}
\frac{g_{1}^{q}(x,p_{T})}{g_{1T}^{\bot q}(x,p_{T})}=\frac{x}{2}\left(
1-\left( \frac{p_{T}}{Mx}\right) ^{2}\right) =\tilde{p}_{1}/M.  \label{m22}
\end{equation}%
Finally, with the use of standard input \cite{lss} on \ $g_{1}^{q}(x)=\Delta
q(x)/2$ we can obtain the curves $g_{1}^{q}(x,p_{T})$ displayed in Fig. \ref%
{ff3}. Let us remark, that the curves change the sign at the point $p_{T}=Mx$%
. This change is due to the term%
\begin{equation}
2x-\xi =x\left( 1-\left( \frac{p_{T}}{Mx}\right) ^{2}\right) =2\tilde{p}%
_{1}/M  \label{m23}
\end{equation}%
in relation (\ref{m20}). This term is proportional to the quark longitudinal
momentum $\tilde{p}_{1}$ in the proton rest frame, which is defined by given 
$x$\ and $p_{T}$. It means, that sign of the $g_{1}^{q}(x,p_{T})$ is
controlled by sign of $\tilde{p}_{1}$. On the other hand, the function $%
g_{1T}^{\bot q}(x,p_{T})$ does not involve term, which changes the sign. The
shape of both functions should be checked by experiment.

To conclude, we presented our recent results on relations between TMDs and
PDFs. The study is in progress, further results will be published
later.\bigskip

{\footnotesize \textbf{Acknowledgements.} A.~E. and O.~T. are supported by
the Grants RFBR 09-02-01149 and 07-02-91557, RF MSE RNP 2.1.1/2512(MIREA)
and (also P.Z.) Votruba-Blokhitsev Programs of JINR. P.~Z. is supported by
the project AV0Z10100502 of the Academy of Sciences of the Czech Republic.
The work was supported in part by DOE contract DE-AC05-06OR23177.

We thank Jacques Soffer and Claude Bourrely for helpful
comments on earlier versions of the manuscript.}


\begin{thebibliography}{99}
\bibitem{tmds} %\bibitem{Collins:2003fm}
J.~C.~Collins, % ``What exactly is a parton density?,''
Acta Phys.\ Polon.\ B \textbf{34}, 3103 (2003); % [arXiv:hep-ph/0304122].\\
%%CITATION = HEP-PH 0304122;%%
%\bibitem{Collins:2007ph}
J.~C.~Collins, T.~C.~Rogers and A.~M.~Stasto, 
%``Fully Unintegrated Parton Correlation Functions and Factorization in Lowest
%Order Hard Scattering,''
Phys.\ Rev.\ D \textbf{77}, 085009 (2008) %[arXiv:0708.2833 [hep-ph]].
%%CITATION = PHRVA,D77,085009;%%
; J.~C.~Collins and F.~Hautmann, Phys. Let. B \textbf{472}, 129 (2000); J.
High Energy Phys. 03 (2001) 016; F.~Hautmann, Phys. Let. B \textbf{655}, 26
(2007).

\bibitem{Mulders:1995dh} P.~J.~Mulders and R.~D.~Tangerman, 
%``The complete tree-level result up to order 1/Q for
%polarized %deep-inelastic leptoproduction,''
Nucl.\ Phys.\ B \textbf{461}, 197 (1996) [Erratum-ibid.\ B \textbf{484}, 538
(1997)] [arXiv:hep-ph/9510301]. %%CITATION = NUPHA,B461,197;%%

\bibitem{Efremov:2009ze} A.~V.~Efremov, P.~Schweitzer, O.~V.~Teryaev and
P.~Zavada, 
%``Transverse momentum dependent distribution functions in a covariant parton
%model approach with quark orbital motion,''
Phys.\ Rev.\ D \textbf{80}, 014021 (2009) [arXiv:0903.3490 [hep-ph]]. 
%%CITATION = PHRVA,D80,014021;%%

\bibitem{Avakian:2009jt} H.~Avakian, A.~V.~Efremov, P.~Schweitzer,
O.~V.~Teryaev, F.~Yuan and P.~Zavada, 
%``Insights on non-perturbative aspects of TMDs from models,''
arXiv:0910.3181 [hep-ph]. %%CITATION = ARXIV:0910.3181;%% 

\bibitem{Zavada:2009sk} P.~Zavada, 
%``Generalized Cahn effect and parton 3D motion in covariant approach,''
arXiv:0908.2316 [hep-ph]. %%CITATION = ARXIV:0908.2316;%%

\bibitem{Zavada:2007ww} P.~Zavada, 
%``Parton distribution functions and quark orbital motion,''
Eur.\ Phys.\ J.\ C \textbf{52}, 121 (2007) [arXiv:0706.2988 [hep-ph]]. 
%%CITATION = EPHJA,C52,121;%%

\bibitem{zav1} P.~Zavada, Phys. Rev. D\textbf{\ 65}, 0054040 (2002).

\bibitem{Martin:2004dh} A.~D.~Martin, R.~G.~Roberts, W.~J.~Stirling and
R.~S.~Thorne, %``Parton distributions incorporating QED contributions,''
Eur.\ Phys.\ J.\ C \textbf{39}, 155 (2005) [arXiv:hep-ph/0411040]. 
%%CITATION = EPHJA,C39,155;%%

\bibitem{Anselmino:2005nn} M.~Anselmino, M.~Boglione, U.~D'Alesio,
A.~Kotzinian, F.~Murgia and A.~Prokudin, 
%``The role of Cahn and Sivers effects in deep inelastic scattering,''
Phys.\ Rev.\ D \textbf{71}, 074006 (2005). %[arXiv:hep-ph/0501196].
%%CITATION = PHRVA,D71,074006;%%

\bibitem{Collins:2005ie} J.~C.~Collins, A.~V.~Efremov, K.~Goeke, S.~Menzel,
A.~Metz and P.~Schweitzer, 
%``Sivers effect in semi-inclusive deeply inelastic scattering,''
Phys.\ Rev.\ D \textbf{73,} (2006) 014021 [arXiv:hep-ph/0509076]. 
%%CITATION = PHRVA,D73,014021;%%

\bibitem{lss} E. Leader, A.V. Sidorov, and D.B. Stamenov, Phys. Rev. D 
\textbf{73}, (2006) 034023.
\end{thebibliography}
\end{document}